\documentclass[prl,twocolumn,showpacs,superscriptaddress,amsmath,amssymb]{revtex4-1}
\usepackage{dcolumn,epsfig,times,color,graphicx,amssymb,amsmath}
 
 \definecolor{Blue}{rgb}{0,0,1}
\definecolor{NavyBlue}{rgb}{0.14,0.14,0.56}
\definecolor{rot}{cmyk}{0,1,1,0}
\newcommand{\e}[1]{\cdot 10^{#1}}  \newcommand{\wn}{\,cm$^{-1}$}
\newcommand{\ea}{\emph{et al.}}  
\newcommand{\bs}{\boldsymbol}

\begin{document}

\title{Splitting  of the Raman $2D$ band of graphene subjected to strain}

\author{M. Mohr}
 \email{marcel@physik.tu-berlin.de}
\affiliation{Institut f\"ur Festk\"orperphysik, Technische Universit\"at Berlin,
Hardenbergstr. 36, 10623 Berlin, Germany}


\author{J. Maultzsch}
\affiliation{Institut f\"ur Festk\"orperphysik, Technische Universit\"at Berlin,
Hardenbergstr. 36, 10623 Berlin, Germany}

\author{C. Thomsen}%
\affiliation{Institut f\"ur Festk\"orperphysik, Technische Universit\"at Berlin,
Hardenbergstr. 36, 10623 Berlin, Germany}


\begin{abstract}
  The Raman $2D$-band -important for the analysis of graphene- shows a splitting for uniaxial strain. The
  splitting depends on the strength and direction of the applied strain and on the polarization of the incident
  and outgoing
  light.  We expand the double-resonance Raman model in order to
  explain the strain direction dependence and the polarization dependence of the splitting. The analysis of this
  splitting gives new insight into the origin of the $2D$-band. Our prediction of the strain
  direction and polarization dependence agrees well with recent experiments.
\end{abstract}


\maketitle

The discovery of graphene in 2004 has led to strong research activities in
the last years\cite{novoselov04}. Graphene has been shown to possess unique material properties.  In
graphene the quantum Hall effect was observed at room temperature
\cite{novoselov05,zhang05}. Because of the specific band structure near the Fermi level, the
electrons behave like massless Dirac fermions and mimic relativistic particles with zero rest mass
and with an effective 'speed of light' $c' \approx 10\e{6}$\,m/s \cite{novoselov05,novoselov07}.
The high electron mobility makes graphene a promising candidate to serve as building block for
micro-electronics.

Detecting and characterizing graphene still is an experimentally difficult task, in particular,
finding its crystallographic orientation. One method to specify graphene layers is Raman
spectroscopy.  This non-destructive method can be used at ambient conditions. It allows to
distinguish between single-layer, bilayer, and multi-layers of graphene\cite{ferrari06}. The quality
of the graphene flakes can be estimated from Raman features\cite{lucchesea10}. Recently, uniaxially
strained graphene has been investigated by Raman spectroscopy\cite{ni08,mohiuddin09,huang09}.  The
strain softens the frequency of the optical phonon branches.  Uniaxial strain also reduces the
hexagonal symmetry of the system. This symmetry breaking leads to new effects, e.g. , the
orientation of graphene can be determined from the polarization dependence of the split
G-band\cite{mohiuddin09,huang09}.  Because of the doping dependence of the G-band-frequency
\cite{yan07} and the higher shift rate of the $2D$-mode the latter is more commonly used to quantify
strain in graphene. The shift of the $2D$-mode is currently under
discussion and usually only a single value for all measurements is given.\cite{mohiuddin09,huang09,tsoukleri09}
However, recent experiments show that the 
shift of the
$2D$-mode in addition depends on the crystallographic orientation and the polarization { of the
  incident and outgoing light during the Raman experiments}.\cite{huang10,privcomm1} In addition,
a splitting of the $2D$-mode has been observed. 
To understand these new results a comprehensive theoretical description is needed to quantify
the shift of the $2D$-mode under strain and that takes into account the polarization and the crystallographic orientation.

Here we { expand} the established model of the double-resonance in graphene.  We explicitly
calculate the resonant Raman cross section which is used to identify the contributing phonons and to
estimate a polarization dependence.  The effect of small strain on the peak positions of the $2D$
mode of graphene is investigated on the level of density-functional theory.  The strain is applied
along arbitrary directions. We are able to explain the splitting, the shift rates, the polarization
effects and the crystallographic orientation dependence.  The observed effects do not solely come
from a movement of the Dirac cones or their deformation, but rather from an orientation dependent
softening of the involved phonon TO branch.  The large magnitude of the splitting can only be
explained with phonons from the branches between the $\Gamma$ and $K$-points that contribute to the
double-resonant $2D$-mode.


\begin{figure}[htb]
\includegraphics[width=0.79\columnwidth]{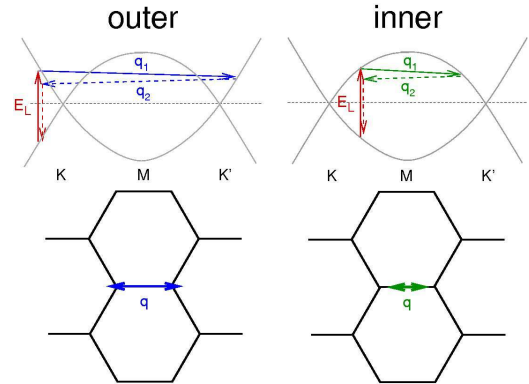} 
\caption{\label{bild:process} Double-resonance mechanism (left: outer process, right: inner process)
  plotted into the band structure scheme of graphene along $KMK$'. Below is the mechanism plotted in
  the Brillouin zone.  }
 \end{figure}

 In Fig.~\ref{bild:process} we show the scheme of the double-resonance principle for the $2D$-mode\cite{thomsen00}.  An
 electron-hole pair is created by an incoming photon with energy $E_l$. The electron or hole is
 scattered inelastically by a phonon with a wavevector $q_1\neq0$ into a real state. From there it
 scattered back inelastically by another phonon $q_2=-q_1$, where it recombines with an energy $E=
 E_l-2\hbar\omega(q_1)$.  The difference to the $2D$-mode is that for the $D$-mode, one inelastic scattering process is
 replaced by an elastic scattering at a defect. It is generally accepted, that the contributing
 phonons stem from the fully-symmetric TO-derived branch\cite{maultzsch04dr}.

 In the following we first use analytical expressions of the electronic and vibrational band
 structure to calculate Raman spectra for different scattering configurations. With these results we identify the
 dominant processes and motivate their extraction from density functional theory (DFT) calculations
 under strain.  The cross  section $|K_{2f,10}|^2$ of a resonant Raman process can be calculated by\cite{cardona82}
\begin{eqnarray}
  \label{cross_sec}
\nonumber  K_{2f,10}=\sum_{a,b,c} & \frac
  {M_{fc}M_{cb}M_{ba}M_{ai}}{(E_l-E_{ai}-i\hbar\gamma)(E_l-\hbar\omega-E_{bi}-i\hbar\gamma)(E_l-\hbar\omega-E_{ci}-i\hbar\gamma)} \\ 
& + \frac
  {M_{fc}M_{cb}M_{ba}M_{ai}}{(E_l-E_{ai}-i\hbar\gamma)(E_l-E_{bi}-i\hbar\gamma)(E_l-\hbar\omega-E_{ci}-i\hbar\gamma)}.
\end{eqnarray}
This is the formula for the $D$-mode. The first term denotes incoming resonance, and the second
outgoing resonance.  $E_{xy}$ is the energy difference between the electronic states $x$ and
$y$. $E_l$ and $E_l-\hbar\omega$ are the energies of the incoming and outgoing photon,
respectively. $M_{xy}$ is the matrix element for the scattering over the intermediate states $x$ and
$y$, and $\gamma$ is the broadening parameter of the electronic transition.

To investigate which processes contribute to the $D$-mode we evaluate Eq.~(\ref{cross_sec}). The sum
is converted into an integral and the matrix elements are taken constant, { although they may
  depend on the electronic wavevector}. The electronic bands are from
a tight-binding approach\cite{wallace47}, where we use the values $\gamma_0=2.8$\,eV and
$\gamma_0\,'=0$\,eV. The TO phonon dispersion around the $K$-point is taken from the analytical
expression derived from recent inelastic X-ray measurements \cite{gruneis09}.

\begin{figure}[htb]
\includegraphics[width=\columnwidth]{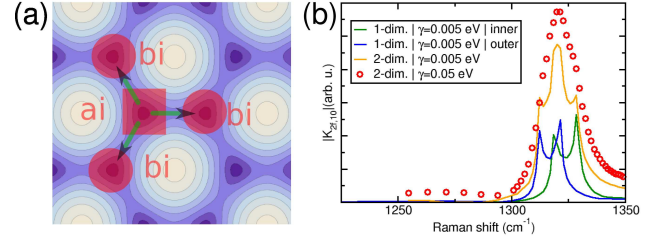} 
\caption{\label{bild:k2f10} 
(a) The two-dimensional integration area plotted in the electronic
  energy contours. The electron absorption happens near ``ai'' and the electron is scattered to
  regions near ``bi''. (b) Raman matrix element $|K_{2f,10}|$ for several different parameter
  sets. Integration is either performed along lines connecting neighboring $K$-points (denoted by
  '1-dim.') or two-dimensional within the highlighted area ('2-dim.'). The similar features in both
  types of integration verifies that contributions from the inner processes are visible in the
 full calculation (two-dimensional integration) as well. At higher values of $\gamma$, single contributions are not
  distinguishable and a single $D$-peak is visible.  
}
 \end{figure}

In a first approximation we integrate along the line that connects two neighboring K-points [see
Fig.~\ref{bild:k2f10}{(a)}]. We have to take special care in this one-dimensional integration as it
leads to erroneous results: in the one-dimensional integrations phonons with $q\approx K$ do not
interfere destructively and give a major contribution\cite{maultzsch04dr}. Still, the one-dimensional
integration gives a good idea for identifying the contributing processes. We find contributions
from phonons from the $K-M$ and $\Gamma -K$ directions.  The Raman matrix element $|K_{2f,10}|$ as a function
of phonon frequency is plotted in Fig.~\ref{bild:k2f10}{(b)}.  Here, a small broadening parameter
of $\gamma=0.005$\,eV was used.  The two peaks of the blue and green plot correspond to incoming and outgoing
resonance.

In a second approximation we now look at the integration in two dimensions.  The integration areas
are highlighted as shaded areas in Fig.~\ref{bild:k2f10}{(a)}, where scattering into all neighboring
$K$-valleys is considered. { The integral is evaluated with the Monte-Carlo method provided in the
Mathematica package (the method 'AdaptiveMonteCarlo' was used with at least $5\times10^6$ points)\cite{mathematica}.}
Now, the $K$-phonon contribution (not shown) is cancelled out due to destructive interference, in
agreement with Ref.~\cite{maultzsch04dr} . Comparing the result in Fig.~\ref{bild:k2f10}{(b)} with
the one-dimensional integration we find similar features. We thus can conclude that phonons from the
$K-M$ direction and phonons from the $\Gamma -K$ direction do contribute to the $D$-mode in
graphene. We will refer to them as outer processes (phonons from $K-M$, see Fig.~\ref{bild:process})
and inner processes (phonons from $\Gamma-K$). Using a larger value of $\gamma=0.05$\,eV, the
distinct features merge and a single broad $D$-band arises [dotted curve in
Fig.~\ref{bild:k2f10}{(b)}].

\begin{figure}[htb]
\includegraphics[width=\columnwidth]{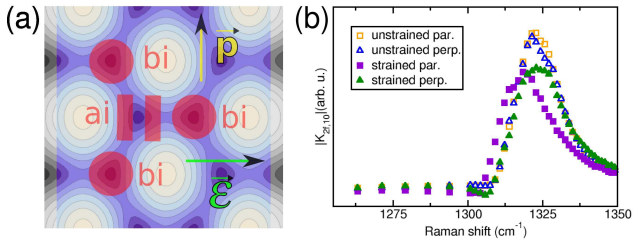}
 \caption{\label{bild:k2f10_EnX} 
(a) Modelling of strain by compressing the band structure. To account for the anisotropic absorption
depending on the polarization, parts of the integration area have been omitted.
(b) Raman matrix element $|K_{2f,10}|$ for strained and unstrained graphene with polarization parallel
and perpendicular to the strain direction. Only when strain is applied, a polarization dependent shift is observed. 
 }
 \end{figure}

 In a perfect graphene flake the $D$-band does not depend on the polarization of the incident light.
 The polarization of the incoming laser affects the absorption in $k$-space. Electrons with a
 wavevector parallel to the polarization of a photon can not be excited during an absorption
 process.
{ We include the effect of polarization by using the $|p\times
 k|^2$-dependence of the absorption, where $p$ denotes the polarization of the incident
 light and $k$ denotes the wavevector of the excited electron\cite{grueneis03}.  This leads to areas in the Brillouin
 zone that give zero contribution to the Raman cross section. } This is illustrated
 in Fig.~\ref{bild:k2f10_EnX}{(a)} for ``vertical'' polarization indicated by the arrow
 $\vec{p}$. As expected and known from experiments, polarization of the incoming light has no
 influence on the $D$-band for unstrained graphene in Fig.~\ref{bild:k2f10_EnX}{(b)}. Now we
 simulate tensile strain, by compressing the band structure in strain-direction. This is indicated
 in Fig.~\ref{bild:k2f10_EnX}{(a)} for strain in the zigzag direction. A strain of 9\% \ was
 used. This is just a proof of concept, as we have not included the phonon softening, and used the
 unstrained phonon dispersion. The resulting Raman cross sections are shown in
 Fig.~\ref{bild:k2f10_EnX}{(b)}. The labels parallel and perpendicular are w.r.t. the strain
 direction. Depending on the polarization, the $D$-band has a different energy shift.

Using DFT calculations of the phonon dispersion and the electronic band structure we now make
quantitative predictions on the peak positions of the 2D band for several scattering configurations.
We have calculated the electronic bands and phonon dispersion of graphene under
strain. These calculations were done with the linear-response DFT code {\sc
  QUANTUM-espresso}\cite{giannozzi09, calc}.  Uniaxial strain was applied \emph{via} the
two-dimensional strain tensor ${\bs\epsilon}=((\epsilon ,0),(0, -\epsilon \nu))$, where $\nu$ is the
Poisson's ratio. Strain in arbitrary directions has been applied through rotating the strain
tensor\cite{strain}.  The strain (tensile and compressive) is applied in several directions between
the zigzag and armchair direction for strain values up to 3\,\%.\cite{calc, strain} However, the
largest differences are found between zigzag and armchair direction. With this data we can now
perform a mapping of the expected $2D$-band frequencies, using the scattering processes shown in
Fig.~\ref{bild:process}. The scattering of the electrons is considered to be elastic for identifying
the contributing phonon's wavevector (this corresponds to horizontal arrows in Fig.~\ref{bild:process}).

\begin{figure}[htb]
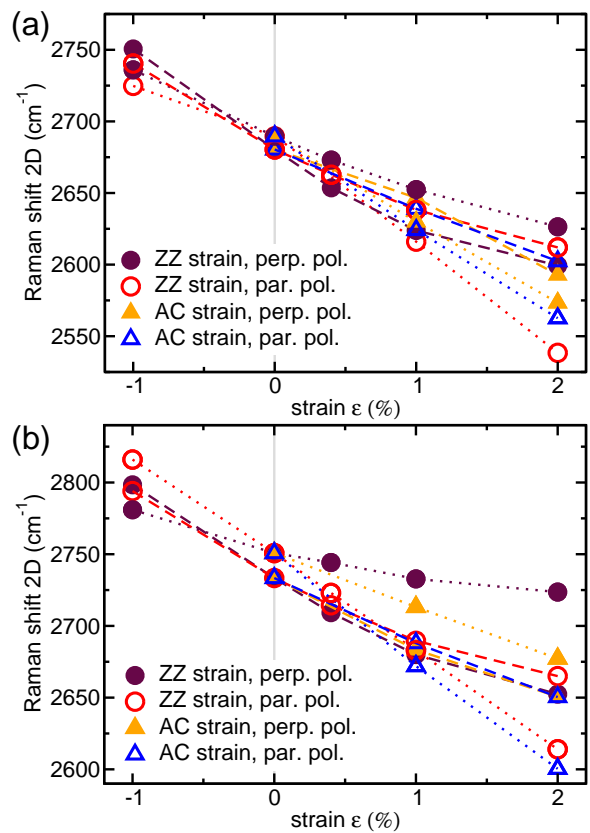

\centering
\includegraphics[width=0.89\columnwidth]{figure4a}
\includegraphics[width=0.89\columnwidth]{figure4b}
 \caption{\label{bild:k2f10_mappingPol} Mapping of the Raman $2D$ contributions for inner and outer
   processes for the laser energies 1.5\,eV (a) and 2.4\,eV (b). Circles (triangles) correspond
   to strain in zigzag (armchair) directions. Open (closed) symbols denote parallel (perpendicular) polarization w.r.t. the strain
   direction.
The data points are connected with lines that denote inner (dotted) and outer (dashed) processes. 
 }
 \end{figure}

\begin{figure}[htb]
\includegraphics[width=\columnwidth]{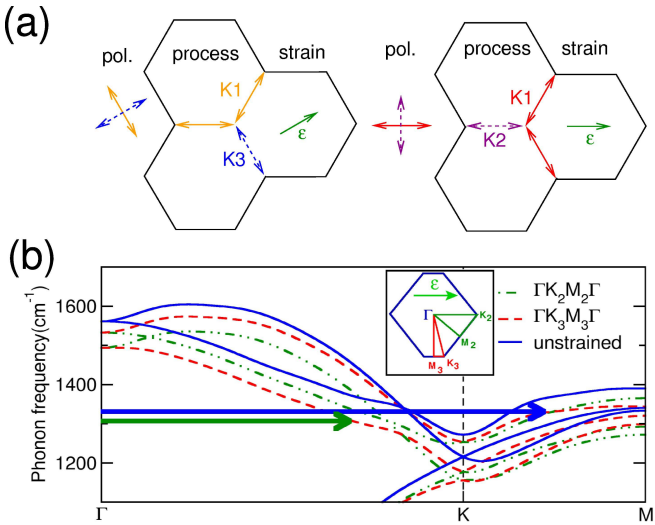}
\caption{\label{bild:Brzone} (a) Dominant processes for perpendicular (solid lines) and parallel
  polarizations (dashed lines) for
  strain in ZZ and AC direction.
  (b) Optical phonon branch along $\Gamma$KM in graphene for unstrained (blue, solid line) and 2\% strain
  in ZZ direction (green, dashed-dotted line and  red, dashed line). The
  corresponding paths in the Brillouin zone are shown in the inset { for an exaggerated strain
    of 30\%. The dark grey (blue) arrow indicates
  the  outer process for a laser energy of 2.4\,eV, the grey (green) arrow indicates the inner process. Note that in absolute units of
  \AA$^{-1}$ the three branches are slightly different due to strain (see inset). In the main figure we
scaled the wavevector axis such that all K and M points are at the same position.}}
 \end{figure}

Depending on which direction in reciprocal space the electron is scattered, different resonance
conditions apply. This is a consequence of the symmetry-breaking induced by the
strain.\cite{mohr09a} In addition, as mentioned above, when the light is polarized, some areas in the
reciprocal space do not have excited electrons that contribute to the $2D$-band. With this argument
the most dominant processes are those that are perpendicular to the polarization of the incoming
light. 
Fig.~\ref{bild:k2f10_mappingPol} now show the obtained $2D$ frequencies for inner and outer processes (as depicted in
Fig.~\ref{bild:process}), for strain in armchair (AC) and zigzag (ZZ) direction, and  for two different laser
energies (1.5\,eV and 2.4 \,eV).  The corresponding processes are illustrated in Fig.~\ref{bild:Brzone}{(a)}. The notation
$K_1$ means along $K_1$, the $K$-points are numbered clockwise, beginning with the top right as
number $1$. 
As can be seen, the
splitting is larger for ZZ strain. Also the polarization dependent shift is larger for ZZ strain.
Both, inner and outer processes lead to a polarization dependent splitting. However, the  inner processes
lead to a  larger splitting.
In Fig.~\ref{bild:Brzone}{(b)} we show the phonon
dispersion along two different paths in the reciprocal space. 
The TO branch between $\Gamma$ and $K$ shows a large splitting depending
on the chosen path. 
The softening is stronger in the direction perpendicular to the strain direction. 
This can be understood looking at the displacement patterns of this mode: 
the displacement is parallel to the strain direction and the involved atomic bonds are weakened due to the strain.
The same argument is responsible for the stronger
softening of the $G^-$-mode at the zone-center\cite{mohiuddin09,huang09,mohr09a}. 
Between $K$ and $M$ the displacement patterns of the TO branch
are not parallel anymore and have lost their TO character and thus are not affected so strongly by the strain direction.


  Biaxial strain was also applied to the system. In this case, the symmetry is conserved, there is no splitting. For the $G$-band we obtain a
  shift rate of 60\,\wn / \%. This is in excellent agreement with the value of 63\,\wn / \% obtained in Ref. \cite{thomsen02} . For the
  $2D$-band we find a shift rate of 135\,\wn /\% for both inner and outer processes. 


  With the above method of frequency mapping we cannot determine the details of intensity and
  lineshape of the $D$ and $2D$-bands.  Simple arguments like looking at the electronic density of
  states do not take into account the phonon density of states.  { Sometimes it is argued that
    the outer processes only contribute to the $2D$-mode because of the trigonal warping effect and
    a higher density of states\cite{kuerti02, narula08}: The trigonal warping effect becomes stronger
    for larger excitation energies, however for optical transition energies $<$2\,eV the equi-energy
    contours are still more or less round.  In addition, the phonon dispersion of the TO-derived
    mode near the $K$-point was a long time believed to show a strong trigonal warping. This was
    recently resolved by Gr\"uneis \ea \ performing inelastic x-ray experiments revealing almost-round
    equi-energy phonons of the TO-branch near the $K$-point\cite{gruneis09} with almost no trigonal
    warping.}

In summary, we have presented an in-depth analysis of the evolution of the $D$ and $2D$-bands in
uniaxially strained graphene.  Depending on polarization and strain direction, different shifts of
the Raman-active $D$ and $2D$-band are predicted.  Strain in zigzag directions leads to a larger
splitting than strain in armchair directions. Our predictions are in excellent agreement with experiments on
strained graphene samples\cite{huang10,privcomm1}.  As the strain-induced shift is used to
determine strain in graphene our results are relevant for the interpretation of experimental data.

We would like to thank H. Yan, M. Huang, J. Hone, and T. Heinz from Columbia University and
O. Frank, K. Papagelis, and C. Galiotis from University of Patras for sharing experimental results
of Refs.~\citenum{huang10,privcomm1}
prior to publication.  JM and CT acknowledge support by the Cluster of Excellence 'Unifying Concepts
in Catalysis' coordinated by the TU Berlin and funded by DFG.


\begin{thebibliography}{10}
\bibitem{calc}%
  \BibitemOpen
  \bibinfo {note} {Calculations were performed with the code {\sc
  QUANTUM-Espresso}\cite{qe}. We used a plane-wave basis set, RRKJ
  pseudopotentials\cite{rappe90} and the generalized gradient approximation in
  the Perdew, Burke and Ernzerhof parameterization for the exchange-correlation
  functional\cite{perdew96}. A Methfessel-Paxton broadening with a width of
  0.02 Ry was used.\cite{methfessel89} The valence electrons were expanded in a
  plane wave basis with an energy cutoff of 60 Ry. A $42\times42\times1$
  sampling grid was used for the integration over the Brillouin zone. The
  dynamical matrices were calculated on a $12\times12\times1$ $q$-grid using
  the implemented linear-response theory. Force constants were obtained via a
  Fourier transformation and interpolated to obtain phonons at arbitrary points
  in the Brillouin zone.}%
  \bibAnnoteFile{Stop}{calc}%
\bibitem{strain}%
  \BibitemOpen
  \bibinfo {note} {For strain in arbitray directions the strain tensor is
  rotated by $\bs\epsilon'=\bs R^{-1}\bs\epsilon\bs R$, where $\bs R$ is the
  rotational matrix. After applying a finite amount of strain we relax the
  coordinates of the basis atoms until forces are below 0.001\,Ry/a.u. and
  minimize the total energy with respect to the Poisson's ratio $\nu$. For
  small strain values we obtain a Poisson's ratio $\nu$=0.164, in excellent
  agreement with experimental tension measurements on pyrolytic graphite that
  yield a value of $\nu$=0.163.\cite{blakslee70}}%
  \bibAnnoteFile{NoStop}{strain}%
\end{thebibliography}
\end{document}